\documentclass[12pt]{article}

\usepackage{amssymb}
\usepackage{epsfig,color}
\usepackage{pstricks,graphicx,epsfig,color,amssymb,amsmath,amscd}
\usepackage{cite}
\usepackage[]{graphicx}
\usepackage{makeidx}

\newcommand{\be}{\begin{eqnarray}}
\newcommand{\ee}{\end{eqnarray}}
\newcommand{\rar}{\rightarrow}

\usepackage[]{caption}
\renewcommand\rho{\varrho}
\date{}

\begin{document}

\begin{titlepage}
\title{ Dynamical mechanism of vacuum energy compensation
%Rates of particle production in $R^2$ gravity and supersymmetry-kind dark matter
}
\author{E.V. Arbuzova$^{a,b}$, A.D. Dolgov$^{b,c}$}

\maketitle
\begin{center}
$^a${Department of Higher Mathematics, Dubna State University, \\Universitetskaya ul. 19, Dubna 141983, Russia}\\
$^b${Department of Physics, Novosibirsk State University, \\Pirogova 2, Novosibirsk 630090, Russia}\\
$^c${Bogolyubov Laboratory of Theoretical Physics, Joint Institute for Nuclear Research,
Joliot-Curie st. 6, Dubna, Moscow region, 141980 Russia}
%$^c${ITEP, Bol. Cheremushkinskaya 25, Moscow 117218, Russia}

%\emailAdd{arbuzova@uni-dubna.ru}
%\emailAdd{dolgov@fe.infn.it}
%\emailAdd{akshalvat01@gmail.com}

\end{center}

%\title{Title}
%\author{}

%\begin{document}

%\maketitle

\begin{abstract}

The model of vacuum energy compensation due to interaction of a scalar field $\phi$ with curvature scalar of the form 
$\beta R \phi^2 f(\phi) $ is proposed. It is shown that with a simple power form of $f(\phi)$ the exponential expansion,
induced by vacuum energy (or what is the same by cosmological constant), is transformed into canonical cosmological 
evolution of the universe dominated by relativistic matter.

\end{abstract}
\thispagestyle{empty}
\end{titlepage}

\section{Introduction}

The problem of the cosmological term or, what is the same, of vacuum energy presents an outstanding challenge to both (quantum)
 field theory and cosmology. Lambda-term was introduced by Einstein at 1917 \cite{Einstein} in futile attempt to make stationary universe. 
 According to Gamow recollection \cite{Gamow}, Einstein considered this idea  
 %  was considered by himself
 as  "biggest blunder" of his life. 
 The generalised by an addition of $\Lambda$-term
 the Hilbert-Einstein action of General Relativity (GR) takes the form:
 %can be generalised by an addition of $\Lambda$-term or
 %a constant density of vacuum energy, $\rho_{vac}$:
 \be
 S =  - \frac{M_{Pl}^2}{16 \pi} \int d^4 x \sqrt{-g} (R +\Lambda) \equiv
 - \frac{M_{Pl}^2}{16 \pi} \int d^4 x \sqrt{-g} R - \rho_{vac}  \int d^4 x \sqrt{-g} ,
 \label{action}
 \ee
 where $g_{\mu\nu}$  is the metric tensor, $g$ is its determinant, 
 $ M_{Pl} = 1.22 \times 10^{19} $ GeV is the Planck mass, related to the Newton gravitational coupling constant as 
 $G_N = 1/M_{Pl}^2 $. %is vacuum energy density. 
  The cosmological constant $\Lambda$ is equivalent to the energy density of vacuum:
 \be
 %\blue
 \rho_{vac} = \frac{M_{Pl}^2 }{16\pi}\, \Lambda. 
 \label{rho-vac-Lam}
 \ee
 The Einstein equations, with Lambda-term included, have the form:
 \be
 \frac{M_{Pl}^2}{8 \pi} \left( R_{\mu\nu} -\frac{1}{2} g_{\mu\nu} R\right) \equiv \frac{M_{Pl}^2}{8 \pi} \, G_{\mu\nu}
 = T_{\mu\nu}^{(matt)} +  T_{\mu\nu}^{(vac)},
 %\rho_{vac} g_{\mu\nu},
 \label{Ein-eq}
 \ee
 where $T_{\mu\nu}^{(matt)}$ is the energy-momentum tensor of matter.  The energy-momentum tensor of vacuum has the form:
 \be
 T_{\mu\nu}^{(vac)} = g_{\mu\nu} \rho_{vac}.
 \label{t-mu-nu-vac}
 \ee
The combination $G_{\mu\nu} = R_{\mu\nu} - g_{\mu\nu} R/2 $ is called the Einstein tensor.

We assume that the space is homogeneous, isotropic, and 3D flat, so the 4D interval  has the form:
\be
 ds^2 \equiv g_{\mu\nu} dx^\mu dx^\nu = dt^2 - a^2(t) d{\bf r}^2 .
\label{ds-2}
\ee
It is convenient to rewrite the Einstein equations in terms of the mixed, upper and lower, indices. 
In particular, using equation $G_0^0 = 3 (\dot a/a)^2 $, which is valid for metric (\ref{ds-2}),
we come to the first Friedman equation:
\be 
\frac{3 H^2 M_{Pl}^2 }{8 \pi} = T^0_0 = \rho_{matt} + \rho_{vac} ,
\label{1-friedman}
\ee
where $H=\dot a /a$ is the Hubble parameter. 

According to theoretical estimates, the magnitude of vacuum energy is by far  larger  than observational bounds on its
possible value.  It is tempting to identify vacuum energy with cosmological dark energy (DE), since they have the same equation of state: 
$P = - \rho$, where $P$ and $\rho $ are respectively the pressure and energy densities.
%of $\rho_{vac}$, that could be identified with dark energy. 
As known, the dark energy makes about 70\% of the total cosmological 
energy density and thus the cosmological density of DE is equal to:
\be
%\rho_{vac} \, {\red \lesssim }\, 
\rho_{DE}\,  { \sim  1} \,{\rm keV/cm}^3 \approx 10^{-47} \,{\rm GeV}^{4}.
\label{rho-DE}
\ee
On the other hand theoretical estimates give either infinitely large  value or, in the case of cancellations of vacuum energies of bosonic 
and fermionic vacuum fluctuations~\cite{vac-Pauli,vac-YaBZ},
the result that could be of the order of the supersymmetry breaking scale:
\be
\rho^{\rm vac}_{\rm SUSY}  \sim m^4_{\rm SUSY} {\sim  10^{55}} \rho_{DE} \,,
\label{rho-sus}
\ee
{where $m_{\rm SUSY} $ is supposed to be of order of $100$ GeV. }
For more detail and extensive list of references see {e.g.} review~\cite{SIB-DAD}.

It is noteworthy that during cosmological evolution, vacuum energy underwent colossal jumps during phase transitions from a symmetrical 
phase to a phase with broken symmetry. 
The problem of vacuum energy becomes especially serious if we consider the structure of the vacuum of quantum 
chromodynamics. As is known, a proton consists of three light quarks: $p = {uud} $, {where 
the  mass of each quark is}, roughly speaking,  5 MeV. Therefore, one should expect that the proton mass should be very small, 
$m_p \approx (15 {\rm MeV} - E_{bind })$,
where $E_{bind}$ is the binding energy of quarks in a proton.
The result obtained is slightly more than 0.01 fraction of the proton mass. 
The missing contribution to the mass comes from the nontrivial properties of the QCD vacuum. Contrary to intuitive expectations, this 
vacuum turns out to be not empty, but filled with a condensate of quark~\cite{cond-q} and gluon fields~\cite{cond-glu}:
\be 
\langle \bar q q  \rangle \neq 0\,, \,\,\,\,\,
 \langle { G_{\mu\nu} G^{\mu\nu} }\rangle  \neq 0\,.
\label{QCD-cnd}
\ee
Energy densities of these condensates are about $\rho_{vac}^{(cond)} \approx $ 1 GeV$^4$. However,
it is argued that quarks inside a proton destroy the condensates and the proton mass would be equal to: 
\be
m_p = 2m_u + m_d - \rho_{vac}^{(cond)} l_p^3 \sim {\rm 1 GeV},
 \label{m-p}
\ee
where $l_p \sim 1/\rm{( GeV)}$ is the proton size. The energy density of the condensate must be negative 
and by 47 orders of magnitude larger than the observed value (\ref{rho-DE}).
Thus there is { experimentally} known contribution to vacuum energy and one has to conclude that
something else, except for quarks and gluons, "lives"  in vacuum and this "something," 
perhaps some new field, compensates $\rho_{vac}$ by {47} orders of magnitude.

There are several suggestions in the literature, 
 where the problem of vacuum energy is discussed. Probably incomplete list of recent works includes 
\cite{Kamenshchik:2018ttr,Barvinsky:2018lyi,Henke:2017cwv,Bengochea:2019daa,Panda:2025clg}.
Somewhat similar approach to that in the presented paper was studied
%In particular, 
in earlier papers~\cite{AD-MK-1,AD-MK-2,AD-MK-3}, where the  singular interaction between
a kinetic term of a scalar field and curvature scalar was considered:
\be
L_{int} = \frac{\partial_\mu \phi  \partial^\mu \phi }{2 R^2}.
\label{L-int-dk}
\ee
In subsequent work~\cite{AD-FU} several different types of the interaction potential between the curvature scalar and a scalar field have
been considered.  However, in each case, the transition to the canonical cosmology dominated by usual matter  was not realised.

The first model of { dynamical} reduction of vacuum energy via interaction of $R$ with a scalar field,
proposed in~\cite{ad-nuff}, involved a massless  scalar field with the action:
\be
S_\phi = \int d^4 x \sqrt{-g} {\cal L} =
\int d^4 x \sqrt{-g} \left[ \frac{1}{2} g^{\mu\nu} \partial_\mu \phi\, \partial_\nu\phi - U(\phi,R) \right] .
\label{s-of-phi}
\ee
Homogeneous field $\phi = \phi (t) $ in FLRW metric (\ref{ds-2}) satisfies the equation of motion: 
%non-minimally related to gravity and satisfying the equation of motion of the form:
%For homogeneous $\phi = \phi(t)$
\be
\ddot \phi + 3H \dot \phi +\partial U/\partial \phi = 0.
\label{ddot-phi-R}
\ee
It is assumed that $\phi$ is non-minimally coupled to gravity with
the potential  $U(\phi,R)$  chosen in the simplest form:
\be
U = \frac{1}{2}\left( \beta R + m^2 \right)\phi^2.
\label{U-of-phi-0}
\ee

It is easy to see that when $\beta R <0 $  and $|\beta R| > |m^2|$
the equation of motion (\ref{ddot-phi-R}) in de Sitter space-time has unstable solutions,
exponentially rising with time, since at constant curvature R the square of the effective mass of the field 
$\phi$ is negative. A similar situation occurs in the Higgs model, when long-wave states with $\phi =0$ turn out to be unstable.

With rising  $\phi$ its influence on cosmological evolution can no longer be neglected and, as 
can be easily verified, the  initial
exponential expansion, $a(t) \sim \exp(H_{vac} t)$,
will asymptotically transform into a power-law one: $\phi \sim t$ and $a(t) \sim t^\kappa$, where $\kappa$
is a certain constant expressed through the model parameters. Thus, the reverse reaction of $\phi$
to the cosmological expansion leads to the transformation of the exponential expansion law into the 
Friedman law, despite the presence of  nonzero vacuum energy.

Among the shortcomings of this simple model  is that the energy-momentum tensor of this field is not 
proportional to the metric tensor, $T_{\mu\nu} \neq \Lambda g_{\mu\nu}$,
and therefore the vacuum energy does not vanish, even asymptotically. The change in the expansion regime is 
achieved due to the 
weakening of the gravitational interaction.  The gravitational coupling constant decreases with time, 
first exponentially, and then as the square of time: $G_N \sim 1/t^2$.
If such a change in $G_N$  took place in the early universe and later somehow stabilised, then this mechanism could 
explain the hierarchy of the gravitational and electroweak scales.

Let us mention in this connection a recent paper~\cite{Montani:2024buy}, where exactly the same statement as in Ref.~\cite{ad-nuff}
was repeated, namely that the effective vanishing of vacuum energy is forced by the time variation of the gravitational coupling constant.

\section{Generalisation of the  model \label{s-mdl-gen}}

In the future we plan to eliminate the shortcomings of the simplest original model with the 
interaction (\ref{U-of-phi-0}) studying  
instead a more general coupling between scalar field $\phi$ and the curvature scalar in the form:
\be
{\cal L}_{F} = \frac{1}{2} \phi^2 \left[\beta R F (R,\phi) +m^2 \right].
\label{L-int}
\ee
In this work  we start with a simpler modification:
\be
{\cal L}_f = \frac{1}{2}\phi^2 \left[\beta R f (\phi) +m^2\right] \equiv 
\frac{1}{2}\left[ \phi^2 m^2 + \beta R Q (\phi) \right],
\label{L-0}
\ee
where we introduced the notation $Q (\phi) = \phi^2 f (\phi)$.

Equation of motion for homogeneous field $\phi$ takes the form:
\be 
g^{\mu\nu}D_\mu D_\nu \phi +m^2 \phi + \frac{1}{2} \beta R \partial_\phi Q =
\ddot\phi + 3 H \dot \phi + m^2 \phi + 
\frac{1}{2} \beta R\, \partial_\phi Q
%\frac{\partial Q(\phi) }{\partial \phi} %Q(\phi)
  = 0,
\label{eq-2-phi}
\ee
where $D_\mu$ is the covariant derivative in metric \eqref{ds-2} and
$\partial_\phi Q = {\partial Q}/{\partial \phi} $.

The energy--momentum tensor of $\phi$ is defined as the variation of the action over the metric tensor: 
\be
T_{\mu\nu} = \frac{2}{\sqrt{-g}}\, \frac{\delta S}{ \delta g^{\mu\nu}}, 
\label{T-mu-nu}
\ee
and correspondingly: 
\be \nonumber
T_{\mu\nu}  &=& (\partial_\mu \phi) (\partial_\nu \phi) 
-(1/2) g_{\mu\nu} \left[ g^{\alpha\beta} (\partial_\alpha\phi)( \partial_\beta \phi) -
 m^2 \phi^2 
 \right] \\
 &&
 - \beta  Q(\phi)  \left(
 R_{\mu\nu} - g_{\mu\nu} R/2 \right) 
+ \beta \left(D_\mu D_\nu - g_{\mu\nu} D^2 \right) Q(\phi).
\label{t-mu-nu-c}
\ee

The covariant derivatives of $Q$  are {equal to:} 
\be
D_\mu Q = (\partial_\phi Q) \partial_\mu \phi,\,\,\, 
D^2 Q = \partial^2_\phi Q \partial_\mu \phi \partial^\mu \phi +\partial_\phi Q D^2 \phi.
\label{dQ}
\ee

Using equation of motion (\ref{eq-2-phi}) we find the trace of the energy-momentum tensor:
\be
T^\nu_\nu =  -(\partial \phi)^2 ( 3\beta \partial_\phi^2 Q +1) + 2 m^2\phi^2 +\beta Q R {+}
3 \beta \left[(\partial_\phi Q)m^2\phi +\beta R (\partial_\phi Q)^2/2 \right] .
\label{trace-1}
\ee

For a special case $Q = \phi^2$ the trace is
\be
T^\nu_{\nu} = -(6\beta+1)( \partial_\mu \phi)( \partial^\mu \phi) + \beta (6\beta+1) R \phi^2 + 2 (1+3 \beta) m^2 \phi ^2. 
\label{trace t-c}
\ee
This is the well known result. 

 Note that for $\beta = -1/6$ and $m = 0$ the trace vanishes.

Now we introduce a more complicated function instead of $f(\phi) $, or better to say instead of $Q(\phi)$,
to cure the problems of a simpler model described above:
\be 
\bar Q(\phi,M_0,k) = \phi^2 (1 + \sigma \phi^2/M_0^2)^k ,
\label{bar-Q}
\ee
{where $M_0$, $k$, and $\sigma$  are some constant parameters to be fixed below. We take $\sigma = \pm 1$. }

The derivatives of $\bar Q$ over $\phi$ are given by:
\be
\partial_\phi {\bar Q } &=&\frac{2k \sigma  \phi^3 (1 + \sigma \phi^2/M_0^2)^{k-1}}{M_0^2} + 2 \phi (1 + \sigma \phi^2/M_0^2)^k,\label{Q'}\\
\partial^2_\phi {\bar Q} &=& \frac{4 ( k-1) k \sigma^2 \phi^4 (1 + \sigma \phi^2/M_0^2)^{k-2}}{M_0^4} + \nonumber \\
&&\frac{ 10 k \sigma \phi^2 (1 + \sigma \phi^2/M_0^2)^{ k-1}}{M_0^2} + 2 (1 + \sigma \phi^2/M_0^2)^k . 
\label{Q'-Q''}
\ee

Taking trace of  Einstein equations (\ref{Ein-eq}) we find the relation:
\be
&&R\left(\beta \bar Q + \frac{3 \beta^2 (\partial_\phi \bar Q )^2}{2} +\frac{M_{Pl}^2}{ 8\pi}  \right) =\nonumber\\
&&( \partial \phi)^2 ( 3\beta \partial_\phi^2 {\bar Q} +1) - 2 m^2\phi^2  
-3 \beta  m^2\phi (\partial_\phi {\bar Q} )  - 4\rho_{vac} - \tilde{T}_\nu^\nu,
\label{R-of-phi}
\ee
where $(\partial \phi)^2 = g^{\mu\nu} (\partial_\mu \phi) (\partial_\nu \phi)$ and $\tilde T^\nu_\nu$ is the trace of the
energy-momentum  tensor of the other kinds of matter.
This equation  allows to express $R$ through $\phi$ and its  first derivative.

\section{Numerical calculations \label{s-numerical}}

There are the following two differential equations {that govern cosmological evolution:}
\be
&&\ddot\phi + 3 H \dot \phi + m^2 \phi + \frac{1}{2} \beta R\, \partial_\phi Q  = 0,\\
&&\dot H + 2 H^2 = - R/6,
\label{sys-eqs}
\ee
where $R$ is a known function of $\phi$ and its { first} derivative given by Eq.~(\ref{R-of-phi}).

To proceed further it is convenient to introduce dimensionless variables:
\be 
&&\tau = t H_0,\, \varphi = \phi/H_0,\, h = H/H_0,\, \frac{d}{dt} = H_0 \frac{d}{d\tau},\nonumber\\
&&R = r H_0^2,\, \rho_{vac} =H_0^4 \lambda,\, \bar Q = H_0^2 q,\,  M_0  = H_0 \mu.
\label{dim-less}
\ee
{Here $H_0$ is a normalisation constant, fixed by the condition %$H_0^2 =  (8\pi/3) (\rho_{vac}/M_{Pl}^2$), 
$H_0^2 =  8\pi\,\rho_{vac}^{(in)}/(3M_{Pl}^2$),
where
$\rho_{vac}^{(in)}$ is the original large vacuum energy.
In what follows we denote derivative over $\tau$ by prime: $df/d\tau \equiv f'$.}

The dimensionless function $q = \bar Q/H_0^2$ is expressed through dimensionless field $\varphi$ as:
%Function $\bar Q = H_0^2\, q$ (\ref{bar-Q}) 
%can be expressed through the dimensionless function $q$ as:
\be
q  = \frac{\bar Q}{H_0^2} = \varphi^2 \left( 1 + \sigma \varphi^2/\mu^2\right)^k
\ee
The derivatives of $\bar Q$ (\ref{Q'}), \eqref{Q'-Q''} turn into:
\be
&&\frac{\partial_\phi {\bar Q}}{ H_0}  \equiv q_1
 = \frac{2 k\sigma \varphi^3 (1 + \sigma {\varphi^2}/{\mu^2})^{k-1}}{\mu^2} + 2 \varphi^2 \left(1 + \sigma {\varphi^2}/{\mu^2}\right)^k,\\ \label{q1}\nonumber\\
&&\partial^2_\phi {\bar Q} \equiv q_2 = \frac{4 ( k-1) k \sigma^2\varphi^4 (1 + \sigma{\varphi^2}/{\mu^2})^{k-2}}{\mu^4} + \nonumber\\
&&\frac{ 10 k \sigma \varphi^2 (1 + \sigma{\varphi^2}/{\mu^2})^{ k-1}}{\mu^2} + 2 \left(1 + \sigma {\varphi^2}/{\mu^2}\right)^k .
\ee

Finally we come to following two differential equations:
\be 
&& h' + 2 h^2 = -\frac{r}{6} \label{dh},\\
&& \varphi'' + 3 h \varphi' + \beta r q_1/2 = 0,
\label{d2phi}
\ee
{where dimensionless curvature $r$ is determined by Eq. \eqref{R-of-phi} and equals to:}
\be 
r= \frac{( 3\beta q_2 +1)  (\varphi')^2 - 2 (m/H_0)^2\varphi^2  - 
3 \beta  (m/H_0)^2 \varphi q_1  -  4\lambda - \tilde{T}/H_0^4 }
{ {3 \beta^2  q_1^2}/{2} {+} M_{Pl}^2/(8 \pi H_0^2) + \beta q } ,
\label{r}
\ee

In what follows we assume that the mass of field $\phi$ is zero, $m = 0$, and that the trace of energy-momentum of other kinds
of matter also vanishes, since it is {supposed to be}
relativistic, $\tilde T_\nu^\nu = 0$, so $r$ turns into
\be
r_0= \frac{( 3\beta q_2 +1)  (\varphi')^2 - 4 \lambda }
{ {3 \beta^2  q_1^2}/{2} {+} M_{Pl}^2/(8 \pi H_0^2) + \beta q } .
\label{r0}
\ee
Equations (\ref{dh}), \eqref{d2phi}  with $r_0$ given by Eq.~(\ref{r0}) are solved numerically taken $\sigma = 1$.
{In Fig.~1 the evolution of dimensionless functions
$h(\tau)$, $\varphi(\tau)$ and $r_0$  are presented at small values of dimensionless time $\tau$, left panel
(a) and large $\tau$, right panel  (b) and for the initial value of dimensionless vacuum energy $\lambda^{(in)} = 10^4$. 
Figure 2 is the same for $\lambda^{(in)} = 10^2$. }
For convenience we multiply all presented quantities by different numerical factors, indicated in figure captions,
to make them of similar magnitude to be depicted in the same figures.

%\newpage
		%\vspace{-0.5cm}
\begin{figure}[htbp]
		%\vspace{-5cm}
		\begin{center}
		\includegraphics[scale=0.5,angle=0]{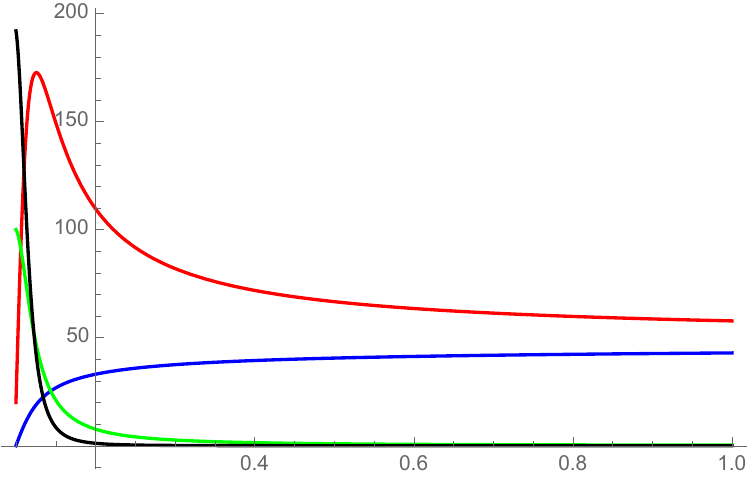} \hspace{2mm}
		\includegraphics[scale=0.5,angle=0]{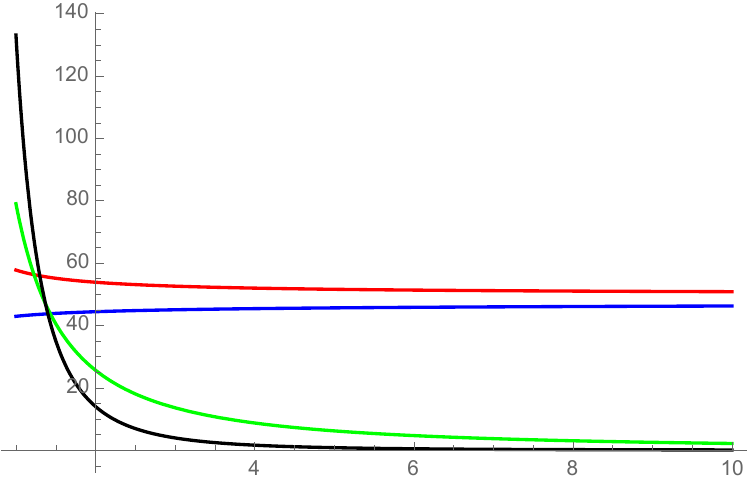}
		\vspace{-3mm}
       \end{center}
	%\vspace{-5cm}
	\caption{{Left panel: small  $\tau$. Evolution of $10^2\tau h$ (red), $10^3\varphi$ (blue),  $10^2\varphi' $ (green) and $(- r_0)$ (black)
	 as functions of time $\tau$. 
	 	Right panel: large  $\tau$. Evolution of $10^2\tau h$ (red), $10^3\varphi$ (blue),  $3 \cdot 10^4\varphi' $ (green) and $(- 10^5 r_0)$ (black)
	 as functions of time $\tau$. 
		 The value of the initial vacuum energy is $\lambda^{(in)} = 10^4$, and $h_{in} =2$, $\varphi_{in} =0$, and $\varphi'_{in} = 1$.
	}}	\label{f-fig1}
\end{figure}

\begin{figure}[htbp]
		%\vspace{-5cm}
		\begin{center}
		\includegraphics[scale=0.5,angle=0]{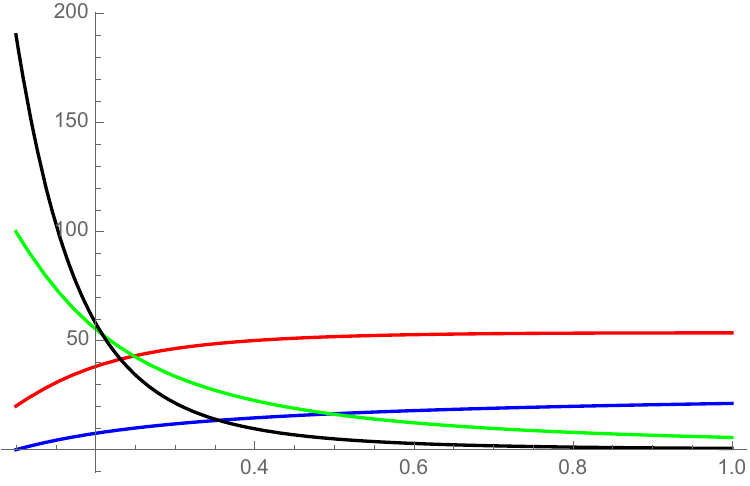}  \hspace{2mm}
		\includegraphics[scale=0.5,angle=0]{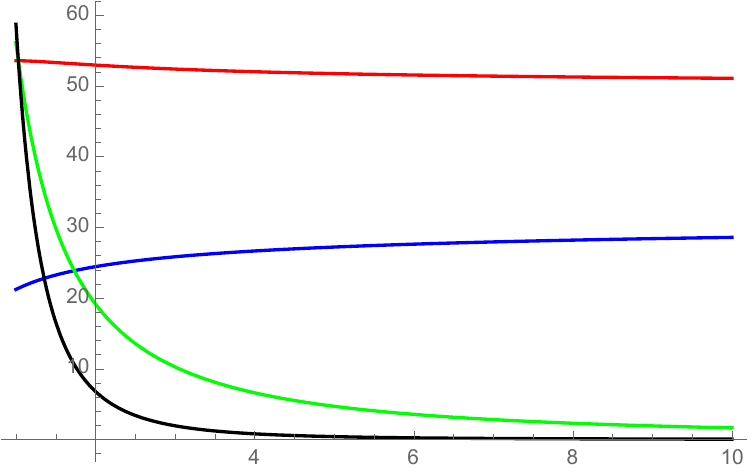}
		\vspace{-3mm}
       \end{center}
	%\vspace{-5cm}
	\caption{{Left panel: small  $\tau$. Evolution of $10^2\tau h$ (red), $10^2\varphi$ (blue),  $10^2\varphi' $ (green) and $(- 10^2 r_0)$ (black)
	 as functions of time $\tau$. 
	 	Right panel: large  $\tau$. Evolution of $10^2\tau h$ (red), $10^2\varphi$ (blue),  $10^3\varphi' $ (green) and $(- 10^4 r_0)$ (black)
	 as functions of time $\tau$. 
		 The value of the initial vacuum energy is $\lambda^{(in)} = 10^2$, and $h_{in} =2$, $\varphi_{in} =0$, and $\varphi'_{in} = 1$.
	}}
	\label{f-fig2}
\end{figure}

Numerical calculations demonstrate that even huge by absolute magnitude initial values of $r_0$ quickly 
{tends} down to zero, thus demonstrating 
that vacuum energy is  compensated. Indeed, a constant value of curvature correponds to a constant value of the Hubble 
parameter and therefore to exponentially expanding de Sitter universe. The impact of field $\phi $ results in asimptotical vanishing of $R$. 
Correspondingly, the de Sitter expansion turns into power law expansion. In other words, vacuum energy is completely compensated and we arrive to 
cosmology dominated by relativistic matter. 

The equations (\ref{dh}), \eqref{d2phi} with $r_0 = 0$ are trivially solved analytically 
leading to the {asymptotic} results:
\be
h(\tau) \rar [2(\tau+\tau_0)]^{-1},\,\,\,  \varphi' \rar C  \tau^{-3/2},\,\,\, \varphi \rar const, 
\label{sol-analyt}
\ee
which pretty well agree with the numerical calculations. Evidently, $H \sim 1/2t$ corresponds to the canonical radiation dominated cosmology.

Let us note that numerical calculations are valid till $\tau \approx  30-40$. At larger $\tau$ instability of the numerical
procedure leads unreasonable results, but at high $\tau$ we fortunately have accurate analytic solutions.

\section{Conclusion \label{s-concl}}

The model suggested in this paper efficiently does the job of elimination any original vacuum energy down to zero and leads to
resulting realistic cosmology governed by relativistic matter. 
According to the relation:
\be
R = -6 (\dot H + 2 H^2), 
\label{R-of-H}
\ee
that is valid in arbitrary homogeneous and isotropic metric, the vanishing of $R$ 
leads to the Hubble parameter evolving as
\be
H = \frac{1}{2(t+t_0)},
\label{h-of-t}
\ee
and correspondingly to the scale factor rising  as $a(t) \sim t^{1/2}$, which is typical the cosmological model
dominated by relativistic matter.

As next steps we need to include into the model non-relativistic matter 
and check if the transition to matter dominated cosmology could be successfully achieved. Another related problem is a possibility
of description of cosmological dark energy in the proposed framework. Presumably it can be realised  by introduction of more 
complicated dependence on curvature scalar, analogously to the known description of dark energy by modified gravity 
through $F(R)$ generalisation of General Relativity. This is supposed to be a matter of future studies.

%\newpage

\section*{Acknowledgments}
This work  was supported by the RSF grant 23-42-00066.

\end{document}